\documentclass[11pt]{article}

\usepackage{amssymb}
\usepackage{color}
\usepackage{graphicx,psfrag}

\begin{document}

\begin{center}
 \textbf{\Large Cinema Fiction vs Physics Reality}\\
 \textbf{\large Ghosts, Vampires and Zombies}\\[5pt]
  Costas J. Efthimiou\footnote{C.J. Efthimiou is a theoretical
 physicist at the University of Central Florida (UCF). He is the advisor
 of the local Campus Freethought Alliance (CFA) chapter which he helped
 to establish at UCF. Address: Department of Physics, UCF, 4000 Central Florida Blvd,
 Orlando, FL 32816. costas@physics.ucf.edu}
  and
  Sohang Gandhi\footnote{S. Gandhi  received his BS on
  physics with honors. Among his many awards, he has been a Goldwater scholar
  and NSF fellow and was selected in the 2006 all-USA third team.
  He has served as the president of the CFA club at UCF. In fall 2006, he
  began his graduate  studies in physics at Cornell University.}
\end{center}

\begin{abstract}
We examine certain features of popular myths regarding ghosts,
vampires and zombies as they appear in film and folklore. We use
physics.
 to illuminate inconsistencies associated with
these myths and to give practical explanation to certain aspects.
\end{abstract}

\section{Introduction}
Perhaps for many, ghosts, vampires, zombies and the like are no
more than Hollywood fantasy.  However, increasingly these movies
have come to reflect the popularity of pseudoscientific beliefs in
the general public. For instance, the movie ``White Noise,"
starring Michael Keaton, is based on the new trend among
paranormalists --- Electronic Voice Phenomena (EVP). The occult
underground in both America and Europe is witnessing  a trendy
rise in vampirism and belief in voodoo zombiefication which is
widespread in many parts of South America and Africa.
Additionally, paranormal depictions in the media, especially TV
and Hollywood motion pictures, have a definite influence on the
way in which people think about paranormal claims
(\cite{Sparks1,Sparks2} and references therein).

In this article we point out inconsistencies associated with the
ghost, vampire and zombie mythologies as portrayed in popular
films and folklore, and give practical explanations to some of
their features. We also use the occasion as an excuse to learn a
little about physics and mathematics.

Of course the paranormalist or occultist could claim that the
Hollywood portrayal is a rather unsophisticated and inaccurate
representation of their beliefs, and thus the discussion we give
hear is moot.  However, if they are to change their definition
each time we raise issue, then all that they are really arguing is
that there exists something out there which may be given the name
`ghost', for instance.  Surely, no skeptic could argue with this.

\section{Ghosts}

\subsection{Sudden Colds}
It has become almost a Hollywood cliche that the entrance of a
ghostly presence be foreshadowed by a sudden and overwhelming
chill (see, for example, ``The Sixth Sense", starring Bruce
Willis). In fact, sharp temperature drops are commonly reported in
association with supposed real-life encounters with ghosts or
poltergeists. This feature of supposed ghost sightings lends
itself naturally to physical explanation.

The famous Haunted Gallery at Hampton Court Palace near London,
UK, is reputedly stalked by the spirit of Catherine Howard, who
was executed on 13 February, 1542, by Henry VIII. Visitors to the
room have described hearing screams and seeing apparitions in the
gallery.  A team of ghost-busting psychologists, led by Dr Richard
Wiseman\footnote{Professor Wiseman's website may be found at
\texttt{www.richardwiseman.com} where details on his research are
presented.} of Hertfordshire University, installed thermal cameras
and air movement detectors in the gallery. About 400 palace
visitors were then quizzed on whether they could feel a
``presence" in the gallery. More than half reported sudden drops
in temperature and some said they sensed a ghostly presence.
Several people claimed to have seen Elizabethan figures.

\begin{figure}[h!]
\begin{center}
\includegraphics[height=5cm]{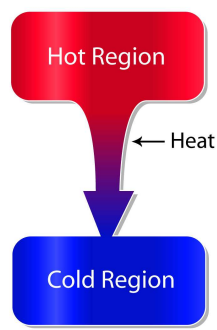}
\end{center}
\caption{Heat always moves from a hotter object to a colder
         object.}
\label{heat}
\end{figure}

Before moving on to an explanation, we will need to outline the
concept of heat. When a `warm' object is placed next to a `cool'
object (see figure \ref{heat}) energy will begin to flow from the
warmer body, causing it to cool, to the cooler body, causing it to
warm. This energy, which is being transferred between the two
objects due to their difference in temperature, is called
\textit{heat}. Note that an object is never said to `possess' any
amount of heat. Heat is only defined through transfer.  For
instance, no matter how high one turns their stove, it never
possesses any degree of heat. In the instance where someone
suddenly touches the stove, however, there is the feeling of heat
--- it is the energy flowing from the stove to that person's hand.

As heat continues to be transferred from the warmer body to the
cooler one in figure \ref{heat}, and the warmer body's temperature
continues to drop while the cooler body's temperature climbs,
there comes a point when the two bodies are at the same
temperature.  At this point heat ceases to flow between the two
object since neither is the hotter one and heat has no definite
direction in which to be transferred.  This condition is called
\textit{thermal equilibrium}.

In our stove example, heat was transferred via \textit{conduction}
--- the exchange of heat through direct contact.  There are two other
modes by which heat may be transferred.  These two modes involve
the exchange of heat by two objects which are separated by some
distance. If these two objects are emersed in a fluid (Earth's
atmosphere for example), then the warmer body may provide heat to
the fluid in its immediate vicinity.  This warmer fluid will then
tend to rise
thus coming in contact with a cooler body above.  There may also
be a lateral current in the fluid, thus allowing the heated fluid
to affect a cooler body to the side. This type of heat transfer,
by an intermediary fluid, is called \textit{convection}.

\begin{figure}[h!]
\begin{center}
\includegraphics[height=4cm]{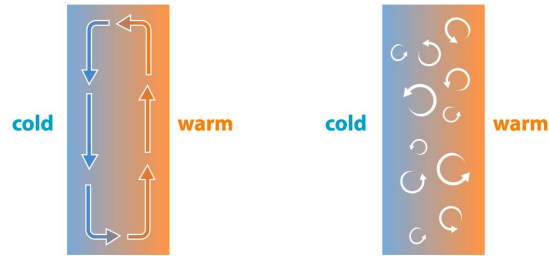}
\end{center}
 \caption{Convection currents in a fluid.}
 \label{convection}
\end{figure}

In figure \ref{convection}(a) we give an example of what is known
as \textit{convection currents}.  Suppose that the right wall is
kept warm and the left wall is kept cool.  Then air in contact
with the right wall will tend to gain heat and rise while air in
contact with the left wall will tend to loose heat and then sink.
The circular flow that then forms is called a convection current.
Air cycles around a loop picking up some heat at the right wall,
dropping it off at the left wall, and then coming back around
again. Actually, the air current pattern will be somewhat more
complicated than what we just described.  There will be all kinds
of smaller cycles and eddies embedded in some complex pattern as
in figure \ref{convection}(b). The overall flow, however, will be
as in figure \ref{convection}(a).

The third mode of heat transfer allows for exchange between two
separated objects even if they are in a total vacuum.  How can two
objects exchange heat if there is no mater in between them?  The
answer is \textit{radiation}.  The thermal energy of a body is
expressed in the `jiggling' of its various constituent particles.
As electrically charged particles within a body jiggle about, they
produce electromagnetic waves. When these waves hit another body,
they cause the particles in that body to jiggle even more than
they were before and thus the body heats up.  Since hotter bodies
produce more of this radiation, there will be more radiation from
the hotter body falling upon the
 cooler body than radiation from the cooler body falling upon the
 hotter body.  Thus, overall, the hotter body will be loosing heat
 while the cooler body will be gaining heat.  We will not be too
 concerned with this particular mechanism for heat exchange here.

Returning to the Haunted Gallery at Hampton Court Palace, Dr
Wiseman's team reported that the experiences could be simply
explained by the gallery's numerous concealed doors. These elderly
exits are far from draught-proof and the combination of air
currents which they let in cause sudden changes in the room's
temperature. In two particular spots, the temperature of the
gallery plummeted by up to $2^\circ$C ($3.6^\circ$F). ``You do,
literally, walk into a column of cold air sometimes," said Dr
Wiseman. ``It's possible that people are misattributing normal
phenomena... If you suddenly feel cold, and you're in a haunted
place, that might bring on a sense of fear and a more scary
experience."  The rumor that `cold spots' are associated with
ghosts seems to be a myth created by the construction of old
building and the vivid imagination of people.

But how could a few degrees drop in temperature explain the
dramatic chills described in so many in ghostly accounts?  First
off, what we sense as cold is not correlated to temperature so
much as the rate at which heat is being transferred from our body
to the environment. For instance, even in a temperate pool, one
feel a very sharp chill when one first enters.  A moderate draft
containing condensed moisture could cause a very sharp sensation
of cold.  Secondly, we are all aware of the `tall-tale' effect.
Memories tend to become distorted and exaggerated.  It is exactly
this reason why scientists tend not to rely on unchecked
eyewitness accounts.

\subsection{The Inconsistency of the Notion of Material-lessness}

Popular myth holds that ghosts are material-less.  For instance in
the movie ``Ghost" (starring Patrick Swayze, Demi Moore, Whoopi
Goldberg), the recently deceased main character tries desperately
to save his former lover from a violent intruder.  His attempts
grant him no avail, as at each lunge he passes right through the
perpetrator. It is interesting, however, that he was able to walk
up the stairs just prior to this.  In fact, this is a common
feature of the ghost myth. Ghosts are held to be able to walk
about as they please, but they pass through walls and any attempt
to pick up an object or affect their environment in any other way
leads to material-less inefficacy
--- unless they are poltergeists, of course!

\begin{figure}[h!]
\begin{center}
\includegraphics[height=4cm]{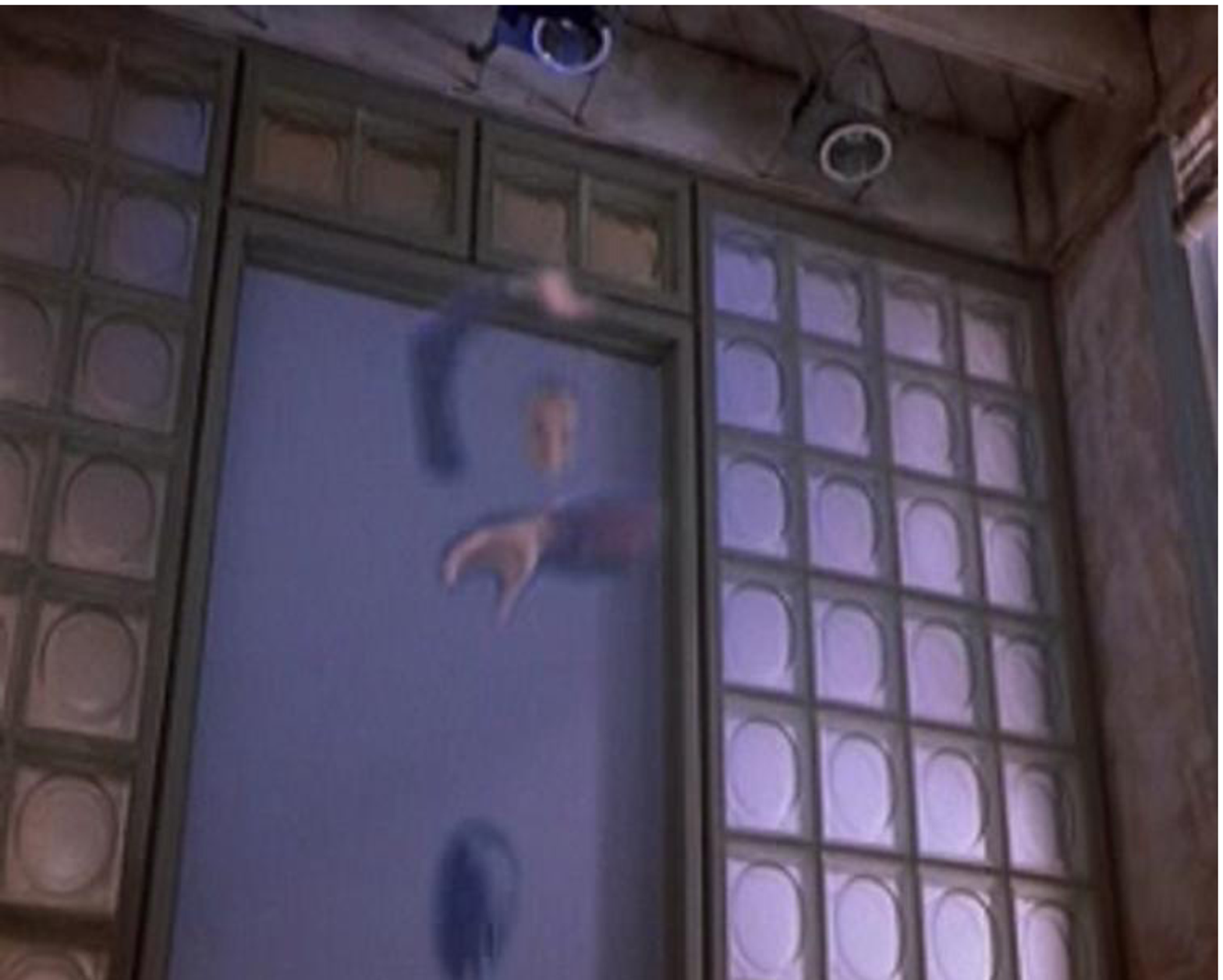}
\hspace{1cm}
\includegraphics[height=4cm]{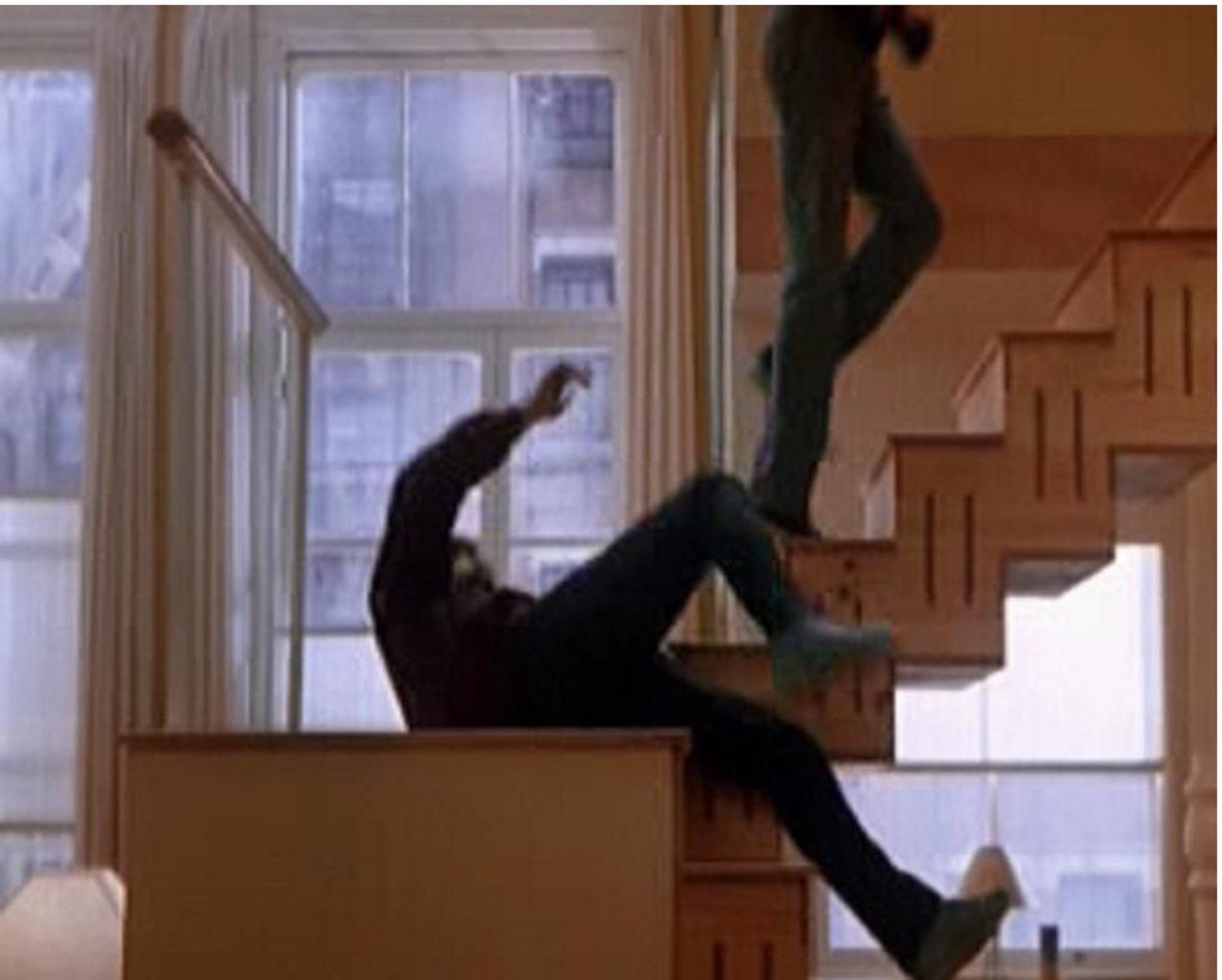}
\end{center}
 \caption{Two stills taken from the movie ``Ghost". I the left still the
          ghost goes through a door. In the right still, the ghost --- which
          follows a burglar in his girlfriend's home ---
          looses his balance as he ascends the staircase and falls on the stairs.}
 \label{ghost}
\end{figure}

Let us examine the process of walking in detail.  Now walking
requires an interaction with the floor and such interactions are
explained by \textit{Newton's Laws of Motion}.  Newton's
\textit{first law} is the law of inertia.  It states that a body
at rest will remain at rest until acted upon by an
\textit{external} force. Therefore, a person cannot start walking
unless a force, applied by some body other than herself, is acting
upon her. But where is the force coming from? The only object in
contact with the person while walking is the floor. So, the force
moving a person during walking is coming from the floor. But how
does the floor know to exert a force when the person wants to
start walking and stop exerting it when the person wants to stand?
Actually, there is no magic here. The person actually tells the
floor. She tells the floor by using Newton's \textit{third law}.

\begin{figure}[h!]
\begin{center}
 \includegraphics[width=7cm]{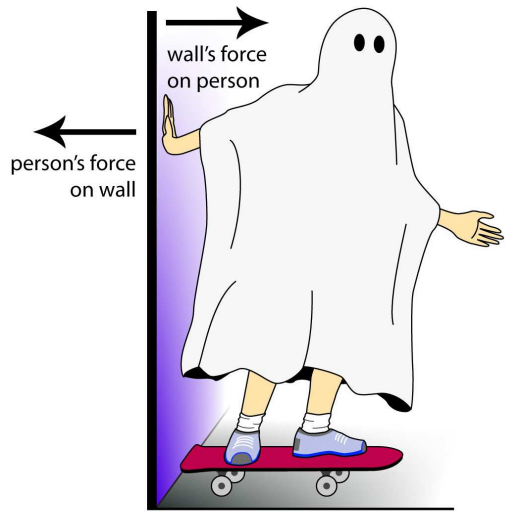}
\end{center}
 \caption{An example of Newton's third law known as action-reaction law.}
 \label{thirdlaw}
\end{figure}

Newton's third law says that if one object exerts a force on
another object, then the second object exerts a force, that is
equal but oppositely directed, on the first object --- hence ``for
each action there is an equal but opposite reaction."  Thus when
the skate-boarder in figure \ref{thirdlaw} pushes on the wall, the
wall pushes right back on her, causing her to accelerate off to
the left.

\begin{figure}[h!]
\begin{center}
 \includegraphics[width=5cm]{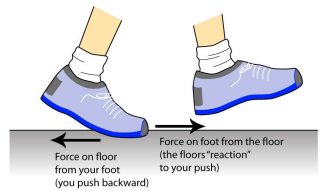}
\end{center}
 \caption{Forces acting on the feet of a person while walking.}
 \label{walking}
\end{figure}

Thus walking goes like this (see figure \ref{walking}):  The
person wanting to do the walking must remain at rest unless a
force acts on her. She gets the floor to apply a force to her by
applying a backward force on the floor with her foot.  She keeps
repeating this action, alternating feet.  The point is that for
the ghost to walk, it must be applying forces to the floor.  Now
the floor is part of the physical universe.  Thus the ghost has an
affect on the physical universe. If this is so, then we can detect
the ghost through physical observation. That is, the depiction of
ghosts walking, contradicts the precept that ghosts are
material-less.

So which is it?  Are ghosts material or material-less?  Maybe they
are only material when it comes to walking.  Well then we must
assume that they can't control this selective material-lessness,
otherwise Patrick Swayze would have saved his girlfriend in
``Ghost." In this case, we could place stress sensors on the floor
and detect a ghost's presence.  Maybe they walk by some other
supernatural means. Well why can't they use this power to
manipulate objects when they want to? Even more, it seems strange
to have a supernatural power that only allows you to get around by
mimicking human ambulation.  This is a very slow and awkward way
of moving about in the scheme of things.  In any case, you'd have
to go to some lengths to make this whole thing consistent.

   Incidentally, the reader may have noticed that we skipped a law in
our discussion. We heard about the first law and the third of
Newton's laws. Newton's second law of motion is that the
acceleration of an object
--- the rate at which it speeds up --- is proportional to the net force
applied, the constant of proportionality being the mass.  We
didn't need the precise statement of this law but, we did make
implicit use of it. The second law implies that the acceleration
of an object will be nonzero (and thus the object will be able to
change its state of motion) only if   a net force is acting on it.
This consistent with our statement `Therefore, a person cannot
start walking unless a force, applied by some body other than
herself, is acting upon her.'

\section{Vampires}
Anyone whose seen John Carpenter's ``Vampires" or  the movie
``Blade" or any of the host of other vampire films is already
quite familiar with how the legend goes.  The vampires need to
feed on human blood. After one has stuck his fangs into your neck
and sucked you dry, you turn into a Vampire yourself and carry on
the blood sucking legacy. The fact of the matter is, if vampires
truly feed with even a tiny fraction of the frequency that they
are depicted to in movies and folklore, then the human race would
have been wiped out quite quickly after the first vampire
appeared.

Let us assume that a vampire need feed only once a month.  This is
certainly a highly conservative assumption given any Hollywood
vampire film.  Now two things happen when a vampire feeds.  The
human population decreases by one and the vampire population
increases by one.  Let us suppose that the first vampire appeared
in 1600 AD.  It doesn't really matter what date we choose for the
first vampire to appear; it has little bearing on our argument. We
list a government website in the references \cite{census} which
provides an estimate of the world population for any given date.
For January 1, 1600 we will accept that the global population was
536,870,911\footnote{It may seem odd to the reader that we have
specified the population with so much precision --- we have a
number in the one-hundred millions and have specified it all the
way down to the `one's place' (...911).
 We chose the particular value for convenience.
The actual estimated population in the 17th century is $562\pm17$
millions.  Beyond mathematical simplification, our choice has
little impact on the argument to follow.  If we were to report any
number in the range of possible values for the population in year
1600, the end result of our calculations below would be
essentially the same.} In our argument, we had at the same time 1
vampire.

We will ignore the human mortality and birth rate for the time
being and only concentrate on the effects of vampire feeding.  On
February 1st, 1600 1 human will have died and a new vampire born.
This gives 2 vampires and $(536,870,911-1)$ humans.  The next
month there are two vampires feeding and thus two humans die and
two new vampires are born.  This gives 4 vampires and
$(536,870,911-3)$ humans.  Now on April 1st, 1600 there are 4
vampires feeding and thus we have 4 human deaths and 4 new
vampires being born.  This gives us 8 vampires and
$(536,870,911-7)$ humans.

By now the reader has probably caught on to the progression.  Each
month the number of vampires doubles so that after $n$ months have
passed there are
\begin{displaymath}
\underbrace{2\times 2 \times \ldots \times 2}_{n~\mathrm{times}} =
2^n
\end{displaymath}
 vampires.  This sort of progression is known in mathematics as a
 \textit{geometric progression} --- more specifically it is a geometric
 progression with ratio 2, since we multiply by 2 at each step.
   A geometric progression increases
 at a very tremendous rate, a fact that will become clear shortly.
 Now all but one of these vampires were once human so
 that the human population is its original population minus the
 number of vampires excluding the original one.  So after
 $n$ months have passed there are
 \begin{displaymath}
536,870,911 - 2^n +1
\end{displaymath}
humans.  The vampire population increases geometrically and the
human population decreases geometrically.

Table \ref{table:vamp} above lists the vampire and human
population at the beginning each month over a 29 month period.
\begin{table}
\begin{center}
\hspace*{-10.5mm}
\begin{tabular}{|c|c|c|c|c|c|}
\hline Month & Vampire Population & Human Population&Month & Vampire Population & Human Population\\
\hline 1 &  1 &  536870911 &  16 & 32768  &     536838144\\
\hline 2 &  2& 536870910 &17 &65536     &  536805376\\
\hline 3 &  4& 536870908 &18 &131072 &536739840\\
\hline 4 &8 &536870904 &19 &262144 &536608768\\
\hline 5  & 16 &536870896 &20 &524288 &536346624\\
\hline 6  &  32 &536870880 &21 &1048576 &535822336\\
\hline 7 &64 &536870848 &22 &2097152 &534773760\\
\hline 8 &  128 &536870784 &23 &4194304 &532676608\\
\hline 9 &256 &536870656 &24 &8388608 &528482304\\
\hline 10 &512 &536870400 &25 &16777216 &520093696\\
\hline 11 &1024 &536869888 &26 &33554432 &503316480\\
\hline 12 &2048 &536868864 &27 &67108864 &469762048\\
\hline 13 &4096 &536866816 &28 &134217728 &402653184\\
\hline 14 &8192 &536862720 &29 &268435456 &268435456\\
\hline 15 &16384 &536854528 &30 &536870912 &0
\\ \hline
\end{tabular}
\caption{Vampire and human population at the beginning of each
month during a 29 month period.} \label{table:vamp}
\end{center}
\end{table}
Note that by month number 30, the table lists a human population
of zero.  We conclude that if the first vampire appeared on
January 1st of 1600 AD, humanity would have been wiped out by June
of 1602, two and half years later.

All this may seem artificial since we ignored other effects on the
human population.  Mortality due to factors other then vampires
would only make the decline in humans more rapid and therefore
strengthen our conclusion.  The only thing that can weaken our
conclusion is the human birth rate.  Note that our vampires have
gone from 1 to 536,870,912 in two and a half year.  To keep up,
the human population would have had to increase by the same
amount. The website \cite{census} mentioned earlier also provides
estimated birth rates for any given time.  If you go to it, you
will notice that the human birth rate never approaches anything
near such a tremendous value. In fact in the long run, for humans
to survive, our population must \textit{at leat} essentially
double each month! This is clearly way beyond the human capacity
of reproduction.

If we factor in the human birthrate into our discussion, we would
find that after a few months, the human birthrate becomes a very
small fraction of the number of deaths due to vampires.  This
means that ignoring this factor has a negligibly small impact on
our conclusion.  In our example, the death of humanity would be
prolonged by only one month.

We conclude that vampires cannot exist, since their existence
contradicts the existence of human beings. Incidently, the logical
proof that we just presented is of a type known as
\textit{reductio ad absurdum}, that is, reduction to the absurd.
Another philosophical principal related to our argument is the
truism given the elaborate title, the \textit{anthropic
principle}.  This states that if something is necessary for human
existence, then it must be true since we do exist.  In the present
case, the nonexistence of vampires is necessary for human
existence.  Apparently, whomever devised the vampire legend had
failed his college algebra and philosophy courses.

\section{Zombies}
The zombie legends portrayed in movies such as ``Dawn of the Dead"
or ``28 Days Later" follow a similar pattern to the vampire
legends. Once you are attacked by zombies, while you may manage to
escape immediate death,  you will eventually die and turn into a
zombie yourself.  Thus this particular type of zombie legend
suffers the same flaw that we pointed out for the vampire legend
previously. We still have some more work to do, however.  There
exists a second sort of zombie legend which pops its head up
throughout the western hemisphere --- the legend of `voodoo
zombiefication'.  This myth is somewhat different from the one
just described in that zombies do not multiply by feeding on
humans but come about by a voodoo hex being placed by a sorcerer
on one of his enemy.  The myth presents an additional problem for
us: one can witness for them self very convincing examples of
zombiefication by traveling to Haiti or any number of other
regions in the world where voodoo is practiced.

We describe the particular case of Wilfrid Doricent\footnote{We
claim no major originality in the presentation of what follows
--- except in collecting the material from the sources and
arranging them as seen. Doricent's case is described in the
documentary \cite{doricent}. The relation between zombies an TTX
was first noticed by the Harvard ethnobotanist Wade Davis in 1982.
He made his hypothesis well known with his book \cite{Davis}.},
an adolescent school boy from a small village in Haiti.  One day
Wilfred had become terribly ill. He was experiencing dramatic
convulsions, his body had swelled terribly and his eyes had turned
yellow.  Eight days latter, Wilfred appeared to have died.  This
was confirmed by not only by the family and family friends present
but also by the local medical doctor who could detect no vital
signs.  Wilfred's body appeared to show bloating due to
rigor-mortis and gave off the foul stench of death and rot. His
body was buried soon thereafter.

\begin{figure}[h!]
\begin{center}
 \includegraphics[width=5cm]{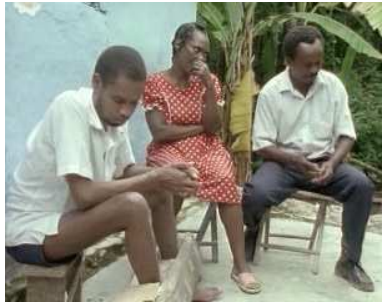}
\end{center}
 \caption{Wilfrid Doricent, the zombie, with his parents.}
 \label{Doricent}
\end{figure}

Some time afterward, the weekly village cock-fight was interrupted
as an incognizant figure appeared.  The villagers were shocked as
they gazed upon the exact likeness of Wilfred.  The arrival was
indeed Wilfred as his family verified by noting scars from old
injuries and other such details. Wilfred, however, had lost his
memory and was unable to speak or comprehend anything.  His family
had to keep him in shackles so that he wouldn't harm himself in
his incoherent state.  It appeared that Wilfred's body had risen
from death leaving his sole in the possession of some voodoo
sorcerer. Word of Wilfred's `zombiefication' spread quickly
throughout the village. It was believed that Wilfred's uncle, a
highly feared voodoo sorcerer who had been engaged in a dispute
over land with Wilfred's family, was the culprit.  Wilfred's uncle
was later charged with zombiefication, a crime in Haiti equivalent
to murder.

Is this truly a case of supernatural magic?  To answer this
question, we turn our attention to a highly toxic substance called
tetrodotoxin (TTX). Bryan Furlow gives an overview \cite{TTX} of
TTX's effects blended with a story from the news:
\begin{center}
\begin{minipage}{13cm}
\small

At first the US federal officers thought they had stumbled upon a
shipment of heroin. The suspicious package they intercepted last
year [2000], en route from Japan to a private address in the US
contained several vials packed with a white crystalline powder.
But on-the-spot tests revealed that it was no narcotic. It took a
while for forensic scientists at the Lawrence Livermore National
Laboratory in California to identify a sample, and what they found
was alarming. The powder turned out to be tetrodotoxin (TTX): one
of the deadliest poisons on Earth.

Gram for gram, TTX is 10,000 times more lethal than cyanide...
This neurotoxin has a terrifying modus operandi---25 minutes after
exposure it begins to paralyze its victims, leaving the brain
fully aware of what's happening. Death usually results, within
hours, from suffocation or heart failure. There is no antidote.
But if luckless patients can hang on for 24 hours, they usually
recover without further complications...

The Livermore team estimated that to extract the 90 milligrams of
TTX discovered by the Feds, you'd need between 45 and 90 kilograms
of puffer fish livers and ovaries---the animal's most deadly
tissues. No one knows what use its intended recipient had in
mind...

 \normalsize
\end{minipage}
\end{center}

\begin{figure}[h!]
\begin{center}
 \includegraphics[width=5cm]{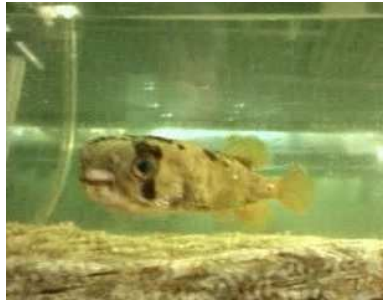}
\end{center}
 \caption{Puffer fish.}
 \label{pufferfish}
\end{figure}

TTX is found in various sea creatures and, in particular, in
various species of puffer fish. Puffer fish are a delicacy in
Japan known as `fugu' where only trained and licensed individuals
prepare it by carefully removing the viscera. Of course, despite
the care taken in preparation, about 200 cases of puffer fish
poisoning are reported per year with a mortality ate 50\%. The
symptoms of the poisoning are as follows \cite{usfda}:
\begin{center}
\begin{minipage}{13cm}
\small The first symptom of intoxication is a slight numbness of
the lips and tongue, appearing between 20 minutes to three hours
after eating poisonous puffer fish. The next symptom is increasing
paraesthesia in the face and extremities, which may be followed by
sensations of lightness or floating. Headache, epigastric pain,
nausea, diarrhea, and/or vomiting may occur. Occasionally, some
reeling or difficulty in walking may occur. The second stage of
the intoxication is increasing paralysis. Many victims are unable
to move; even sitting may be difficult. There is increasing
respiratory distress. Speech is affected, and the victim usually
exhibits dyspnea, cyanosis, and hypotension. Paralysis increases
and convulsions, mental impairment, and cardiac arrhythmia may
occur. The victim, although completely paralyzed, may be conscious
and in some cases completely lucid until shortly before death.
Death usually occurs within 4 to 6 hours, with a known range of
about 20 minutes to 8 hours.
 \normalsize
\end{minipage}
\end{center}

Sometimes however, a victim pronounced dead, is lucky enough to
wake up just before his funeral and report to his bewildered
family that he was fully conscious and aware of his surroundings
the entire ordeal. Therefore, TTX has the unusual characteristic
that if a nonlethal dose is given, the brain will remain
completely unaffected. If just the right dose is given, the toxin
will mimic death in the victim, whose vitals will slow to an
immeasurable state, and whose body will show signs of rigor-mortis
and produce the odor of rot. Getting such a precise dose would be
rare for the case of fugu poisoning, but can easily be caused
deliberately by a voodoo sorcerer, say, who could slip the dose
into someone's food or drink.

\begin{figure}[h!]
\begin{center}
 \includegraphics[width=5cm]{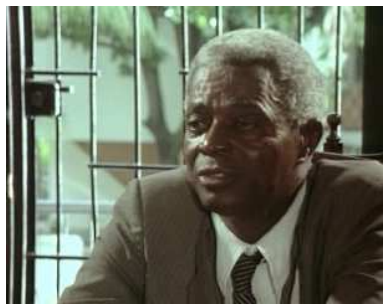}
\end{center}
 \caption{Fr\`ere Dodo, ex-voodoo priest, confirms that the recipe used to make the drug for
          zombiefication includes a power derived from the puffer fish.}
 \label{priest}
\end{figure}

The secrets of zombiefication are closely guarded by voodoo
sorcerers.  However, Fr\`{e}re Dodo, a once highly feared voodoo
sorcerer who is now an Evangelical  preacher and firm denouncer of
the voodoo faith, has revealed the process.  It turns out that
zombiefication is accomplished by slipping the victim a potion
whose main ingredient is powder derived from the liver of a
species of puffer fish native to Haitian waters.

Well, we now have an explanation for how Wilfred could have been
made to seem dead, even under the examination of a doctor.
However, we have already said that the TTX paralysis was unlikely
to have affected his brain.
 How does one account for Wilfred's comatose mental state? The answer is oxygen
deprivation. Wilfred was buried in a coffin in which relatively
little air could have been trapped. Wilfred's story probably goes
something like this: Slowly, the air in Wilfred's coffin began to
run out so that by the time he snapped out his TTX-induced
paralysis, he had already suffered some degree of brain damage.
At this point his survival instincts kicked in and he managed to
dig himself out of his grave --- graves tend to be dug shallow in
Haiti. He probably wondered around for some time before ending up
back the village.

\begin{figure}[h!]
\begin{center}
 \includegraphics[width=5cm]{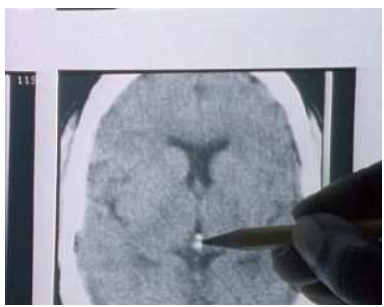}
\end{center}
 \caption{A picture from the brain scan of W. Doricent.}
 \label{brainscan}
\end{figure}

Neuropsychiatrist Dr. Roger Mallory, of the Haitian Medical
Society, conducted a scan of zombiefied Wilfred's brain. Although
the results were not as definite as had been hoped for, he and his
colleagues found brain damage consistent with oxygen starvation.
It would seem that zobiefication is nothing more then a skillful
act of poisoning. The bodily functions of the poisoned person
suspend so that he appears dead. After he is buried alive, lack of
oxygen damages the brain. If the person is unburied before he
really dies from suffocation, he will appear as a soulless
creature (`zombie') as he has lost what makes him human: the
thinking process of the brain\footnote{We must observe that Davis'
link between zombiefication and TTX has been viewed with
skepticism. Unfortunately, there has been no direct proof along
this direction and, therefore, Davis' claim remains an intriguing
hypothesis. There are only a few scientific studies on zombies (at
least to the best of our knowledge). One interesting study is that
of the anthropologist Rolland Littlewood and Dr Chavannes Douyon
\cite{LD} who looked at three alleged zombie cases. In all cases
they were able to find medical explanations.}.

\section{Conclusion}
We have examined the science behind three of the most popular
pseudoscientific beliefs encountered in Hollywood movies. For two
of them --- the idea of ghosts and vampires --- we have shown that
they are inconsistent and contradictory to  simple facts. For one
of them
--- the idea of zombies --- we have made no attempt to deny that it
relies on real cases. However, we have reviewed evidence showing
that the concept is a misrepresentation of simple criminal acts.

Wide spread belief in such concepts, we feel, is an indication of
a lack of critical thinking skills in the general population. With
simple elementary arguments one can easily discredit the validity
of such claims. We thus finish with the following quote by Carl
Sagan \cite{Sagan}:
\begin{quote}
{\small Both  Barnum and  H.L. Mencken are said to have made the
depressing observation that no one ever lost money by
underestimating the intelligence of the American Public. The
remark has worldwide application. But the lack is not
intelligence, which is in plentiful supply; rather, the scarce
commodity is systematic training in critical thinking.}
\end{quote}

\newpage

\section*{Addendum}
After its initial release
the present article received great attention by the media and the
public. In addition, letters from readers were received after the
publication of the article in the July/August 2008 issue of
\emph{Skeptical Inquirer}.

The authors would like to thank all readers who took the time to
send their comments although  we were surprised (and partly
disappointed) to see that the majority of the readers
misunderstood the goal of the article. Since most of the comments
have considerable overlap, we thought we'd summarize our
explanations in one reply.

We would like to point out that our article was not about definite
proofs.  We challenge the reader to devise an absolute,
mathematical/logical proof that a given supernatural occurrence
does not exist.  Our prediction is that the reader will fail.  In
particular there is no universally agreed upon
mathematical/logical definition of the various apparitions
considered in our article nor of the manner in which they
operate/behave (e.g. whether vampires deliberately control which
of their victims turns into a vampire, etc; cf the vast majority
of the received comments).  However, an inability to present a
definite proof does not imply an inability to discover logical and
scientific inconsistencies or other flaws in a claim.

Our article was intended merely as an entertaining vehicle of
education aiming at stirring critical thinking.  Our goal was to
remind readers that  (a) Pseudoscientific and paranormal ideas
barely make sense when elementary logic and science is applied;
(b) When---and if---there is an element of truth, it is highly
distorted and hidden behind elaborate myths; (c) To teach readers
a little about science and remind them to make probabilistic
assessments of various claims using reason.

Many letters state that the authors have `missed' essential ideas
in the `vampirization' of humans that invalidate the calculation
and, thus, the final conclusion. We would like to assure the
readers that we are familiar with all the variations of the myth.
After all, how could we miss them if Hollywood and novel writers
bombard us with them daily? We know about Buffy, Angel, Blade and
the other vampire hunters. We have read about multiple bites,
drained bodies, transfer of blood and other protocols necessary to
create new vampires. However, none of them can change the final
conclusion of the `impossibility of the existence of vampires'
albeit the line of reasoning might have to be modified slightly
(or radically if the premises are changed considerably). No matter
what assumptions are required, one can create a corresponding
mathematical model. The authors intentionally simplified the
assumptions and avoided sophisticated mathematical models lest the
article be inaccessible to some---depending on education.  By
introducing dynamical systems, one can construct highly
sophisticated models for the vampire population versus the human
population. For example, in one of the simplest models, known as
the \emph{prey-predator model}, the two populations fluctuate
periodically. (For a simple presentation see \cite{Stewart}.) If
this model is used, the human population never disappears but it
fluctuates between a maximum and a minimum value. One can
immediately see now an argument against the existence of vampires:
the human population has not fluctuated and even more it has kept
(and keeps) increasing exponentially.  One can start with this
model and add additional features (such as vampires slayers,
vampire diseases, accidental exposure to sunlight, vampire babies,
etc) but the final outcome will not change: vampires cannot exist
since the model would predict a human population curve different
from the actual one.

Also, the reader should note that this discussion is based in
elementary physics and mathematics. We never discussed the social
implications. Imagine what it would mean if every so often an
exsanguinated human corpse was found (even if only one vampire
existed). Wouldn't this be \emph{the} headline in the news?
Unless, of course, all governments have conspired to keep these
incidents secret...

Other messages to the authors pointed out similar omissions/holes
in the arguments of ghosts and tried to affirm the existence of
ghosts based on faith or incorrect physics explanations.
Unfortunately quantum mechanics and exotic matter cannot give more
substance to ghosts and faith cannot be used as a substitute for
proof.  For reasons of limited space and time we bypass a rebuttal
of each of the (incorrect) attempts to use physics to make the
concept of ghosts consistent.

Many Hollywood movies have some ideas that are consistent with
science ideas but the majority of movies greatly offend
mathematical and scientific laws. Our reference to the movies was
a motivational tool, mainly. The authors would enjoy a consistent
script although research on the issue  indicates that Hollywood
has a negative impact on scientific literacy in the public.
Ignoring misrepresentation of particular situations and even
ignoring series like \emph{SciFi Investigates},
\emph{Ghusthunters}, etc which involve people untrained in the
scientific method of investigation, Hollywood's current trend is
to promote the supernatural over logic and scientific inquiry.
Science is irrelevant, it means trouble, is reflexively
close-minded, and only laypersons will find the true solution.
Fortunately, a few people have tried to reverse the unchallenged
way Hollywood presents its ideas. There have been some excellent
books based on Hollywood products that explain science in an
entertaining way. Among our favorites are the classic book by
Lawrence Krauss \emph{The Physics of Star Trek} and James
Kakalios's \emph{The Physics of Superheroes}. One of us (C.E.) is
involved in a more extensive project nicknamed \emph{Physics in
Films} \cite{PiF} that uses Hollywood movies as a vehicle of
education to increase the scientific literacy and the quantitative
fluency of the public. The projects also attempts to reverse the
unchallenged way that Hollywood promotes its ideas.

\end{document}